\documentstyle[12pt,epsfig]{article}

\catcode`\@=11
\long\def\@makefntext#1{
\protect\noindent \hbox to 3.2pt {\hskip-.9pt
$^{{\ninerm\@thefnmark}}$\hfil}#1\hfill}                

\def\@makefnmark{\hbox to 0pt{$^{\@thefnmark}$\hss}}  

\def\ps@myheadings{\let\@mkboth\@gobbletwo
\def\@oddhead{\hbox{}
\rightmark\hfil\ninerm\thepage}
\def\@oddfoot{}\def\@evenhead{\ninerm\thepage\hfil
\leftmark\hbox{}}\def\@evenfoot{}
\def\sectionmark##1{}\def\subsectionmark##1{}}

\renewcommand{\thefootnote}{\fnsymbol{footnote}}

\newcounter{sectionc}\newcounter{subsectionc}\newcounter{subsubsectionc}
\renewcommand{\section}[1] {\vspace*{0.6cm}\addtocounter{sectionc}{1}
\setcounter{subsectionc}{0}\setcounter{subsubsectionc}{0}\noindent
        {\normalsize\bf\thesectionc. #1}\par\vspace*{0.4cm}}
\renewcommand{\subsection}[1] {\vspace*{0.6cm}\addtocounter{subsectionc}{1}
        \setcounter{subsubsectionc}{0}\noindent
        {\normalsize\it\thesectionc.\thesubsectionc. #1}\par\vspace*{0.4cm}}
\renewcommand{\subsubsection}[1]
{\vspace*{0.6cm}\addtocounter{subsubsectionc}{1}
        \noindent
{\normalsize\rm\thesectionc.\thesubsectionc.\thesubsubsectionc.
        #1}\par\vspace*{0.4cm}}

\newcounter{appendixc}
\newcounter{subappendixc}[appendixc]
\newcounter{subsubappendixc}[subappendixc]

\renewcommand{\appendix}[1] {\vspace*{0.6cm}
        \refstepcounter{appendixc}
        \setcounter{figure}{0}
        \setcounter{table}{0}
        \setcounter{equation}{0}
        \renewcommand{\thefigure}{\Alph{appendixc}.\arabic{figure}}
        \renewcommand{\thetable}{\Alph{appendixc}.\arabic{table}}
        \renewcommand{\theappendixc}{\Alph{appendixc}}
        \renewcommand{\theequation}{\Alph{appendixc}.\arabic{equation}}
        \noindent{\bf Appendix \theappendixc #1}\par\vspace*{0.4cm}}

\def\abstracts#1{{

\centering{\begin{minipage}{12.2truecm}\footnotesize\baselineskip=12pt\noindent
        \centerline{\footnotesize ABSTRACT}\vspace*{0.3cm}
        \parindent=0pt #1
        \end{minipage}}\par}}

\renewenvironment{thebibliography}[1]
        {\begin{list}{\arabic{enumi}.}
        {\usecounter{enumi}\setlength{\parsep}{0pt}
\setlength{\leftmargin 1.25cm}{\rightmargin 0pt}
         \setlength{\itemsep}{0pt} \settowidth
        {\labelwidth}{#1.}\sloppy}}{\end{list}}

\newcounter{itemlistc}
\newcounter{romanlistc}
\newcounter{alphlistc}
\newcounter{arabiclistc}

\newcommand{\fcaption}[1]{
        \refstepcounter{figure}
        \setbox\@tempboxa = \hbox{\footnotesize Fig.~\thefigure. #1}
        \ifdim \wd\@tempboxa > 6in
           {\begin{center}
        \parbox{6in}{\footnotesize\baselineskip=12pt Fig.~\thefigure. #1}
            \end{center}}
        \else
             {\begin{center}
             {\footnotesize Fig.~\thefigure. #1}
              \end{center}}
        \fi}

\newcommand{\tcaption}[1]{
        \refstepcounter{table}
        \setbox\@tempboxa = \hbox{\footnotesize Table~\thetable. #1}
        \ifdim \wd\@tempboxa > 6in
           {\begin{center}
        \parbox{6in}{\footnotesize\baselineskip=12pt Table~\thetable. #1}
            \end{center}}
        \else
             {\begin{center}
             {\footnotesize Table~\thetable. #1}
              \end{center}}
        \fi}

\def\@citex[#1]#2{\if@filesw\immediate\write\@auxout
        {\string\citation{#2}}\fi
\def\@citea{}\@cite{\@for\@citeb:=#2\do
        {\@citea\def\@citea{,}\@ifundefined
        {b@\@citeb}{{\bf ?}\@warning
        {Citation `\@citeb' on page \thepage \space undefined}}
        {\csname b@\@citeb\endcsname}}}{#1}}

\newif\if@cghi
\def\cite{\@cghitrue\@ifnextchar [{\@tempswatrue
        \@citex}{\@tempswafalse\@citex[]}}
\def\citelow{\@cghifalse\@ifnextchar [{\@tempswatrue
        \@citex}{\@tempswafalse\@citex[]}}
\def\@cite#1#2{{$\null^{#1}$\if@tempswa\typeout
        {IJCGA warning: optional citation argument
        ignored: `#2'} \fi}}

 1
 1
 1

\font\ninerm=cmr9

\begin{document}

\vspace*{-2.6cm}
\begin{flushright}
KA-THEP-5-1995\\
March 1995
\end {flushright}

\vspace{2.0cm}

\centerline{\normalsize\bf Z BOSON OBSERVABLES IN THE MSSM\footnote{
      To appear in the Proceedings of the Ringberg Workshop
      ``Perspectives for Electroweak Interactions in $e^+$ $e^-$
      Collisions'', February 5-8, 1995.}}

\baselineskip=22pt
\vspace{0.4cm}

\centerline{\footnotesize A. DABELSTEIN\footnote{E-Mail:
add@dmumpiwh.bitnet, wom@dmumpiwh.bitnet} \ ,
W. HOLLIK, W. M\"OSLE$^\dagger$}
\vspace{0.3cm}

\baselineskip=13pt
\centerline{\footnotesize\it Institut f\"ur Theoretische Physik,
 Universit\"at Karlsruhe, Kaiserstr. 12}
\baselineskip=12pt
\centerline{\footnotesize\it D-76128 Karlsruhe, Germany}
\centerline{\footnotesize E-mail: add@dmumpiwh.bitnet}
\vspace*{0.3cm}

\vspace*{0.9cm}
\abstracts{
 A combined analysis for $Z$ boson observables and the quantity $\Delta r$
 within the framework of the Minimal Supersymmetric Standard Model is
presented.
 Contributions from supersymmetric particles are discussed and upper and
 lower bounds for the MSSM predictions are given.}

\normalsize\baselineskip=15pt
\setcounter{footnote}{0}
\renewcommand{\thefootnote}{\alph{footnote}}
\section{Introduction}
Precision data at LEP have reached a level of accuracy where the
electroweak theory can be tested at its quantum structure.
The predictions of the Standard Model for the precision observables at the
$Z$ resonance have been calculated with full one-loop and
leading higher order contributions\cite{bhp,Chet/Kw}.
With two exceptions, all precision observables, measured at the level of
a few per mille or even better,
agree with the Standard Model predictions
within $1 \sigma$ \cite{Lep1}.
 The top quark mass $m_t = 180 \pm 12$ GeV as
the weighted average of the CDF and D\O \  measurements\cite{top} is in
good agreement with the Standard Model implication
from the precision data. The only mismatch is found in
the ratios of the hadronic partial decay widths $Z
\rightarrow c \bar{c} \, (b \bar{b})$ to the total hadronic width :
$$
R_c = \frac{\Gamma_c}{\Gamma_{had}} \ , \
R_b = \frac{\Gamma_b}{\Gamma_{had}} \ ,
$$
which
deviate by $\approx 2 \sigma$ from the recent LEP data\cite{Lep1}.
\smallskip

It is not unlikely that the Standard Model around the $Z$ scale is the
effective low energy limit of a more comprehensive theory of the
GUT class. Within these ideas the extension of the Standard Model to the
Minimal Supersymmetric Standard Model (MSSM)\cite{hunter} is of particular
theoretical interest, allowing the unification of gauge couplings at
the GUT scale $\cal{O} \rm (10^{15}$ GeV$)$\cite{boer}.

\smallskip
At present the MSSM is the most predictive
framework for physics beyond the Standard Model.
The superpartners appear in the electroweak radiative corrections and thus
enter the theoretical predictions for the precision observables
with the property that they decouple in the heavy mass limit.
Precision measurements therefore provide a tool for indirect tests of the
MSSM and for probing the possible mass range of new particles beyond the
present limits from negative direct searches.

\smallskip
As an extension of the Standard Model, the general 2-Higgs doublet has been
discussed
for $Z$ boson decays\cite{Denner} and implications from electroweak
precision data are presented in ref.\cite{Cornet}, also for a
supersymmetric Higgs sector under the assumption of decoupling genuine
SUSY particles.
Recently, supersymmetric
contributions for individual $Z$ boson decay widths and ratios
$R_b$, $R_c$ were
presented in refs\cite{Boulware,Wells,Garcia1,carena,Bhatta}.

\smallskip
In this article we describe
the results of a complete MSSM 1-loop calculation for a
combined discussion of all $Z$ observables and $M_W$. The
$Z$ boson observables together with the quantity $\Delta r$ (or $M_W$)
 depend sizeably on
the masses and parameters of the virtual standard  and supersymmetric
 particles. For each observable
the effects from the parameters in the various relevant MSSM sectors
are discussed and comparisons with the minimal Standard Model are
performed.
Considering the $Z$ boson observables and $\Delta r$
globally as functions of the MSSM parameter set, the following conclusions can
be observed:
\begin{itemize}
 \item[$\bullet$] the MSSM can improve the agreement of the
                individually predicted observables with the
                experimental data. It is, however, difficult to remove the
                discrepancies from all observables simultaneously.
  \item[$\bullet$] a $\chi^2$ fit for all input parameters gives a prefered
                set  of
                MSSM parameters. This numerical analysis yields constraints on
                the supersymmetric particle masses and parameters. A prediction
                for a supersymmetric particle spectrum can not be achieved
                without additional assumptions,
                since
                the precision observables are only weakly sensitive to equal
                supersymmetric masses $\ge 500$ GeV\cite{dab/hollik}.
\end{itemize}
In the following, the considered $Z$ boson observables and their notation are
introduced. A brief description of the one-loop and higher order corrections
of the precision observables is presented and the parametrization of the MSSM
is given. The discussion compares the MSSM predictions with the
experimental data and presents constraints on the supersymmetric parameter
space.

\section{$Z^0$ boson on-resonance observables}

At the $Z$ boson resonance two classes of precision observables are available:
\begin{itemize}
\item[a)] inclusive quantities:
  \begin{itemize}
   \item[$\bullet$] the partial leptonic and hadronic decay width $\Gamma_{f
                  \bar{f}}$,
   \item[$\bullet$] the total decay width $\Gamma_Z$,
   \item[$\bullet$] the hadronic peak cross section $\sigma_h$,
   \item[$\bullet$] the ratio of the hadronic to the electronic decay
                  width of the $Z$ boson: $R_h$,
   \item[$\bullet$] the ratio of the partial decay width for $Z\rightarrow
c\bar{c} \,
                  ( b \bar{b} )$ to the hadronic width,
                  $R_c$, $R_b$.
  \end{itemize}
\item[b)] asymmetries and the corresponding mixing angles:
  \begin{itemize}
   \item[$\bullet$] the \it forward-backward \rm asymmetries $A_{FB}^f$,
   \item[$\bullet$] the \it left-right \rm  asymmetries $A_{LR}^f$,
   \item[$\bullet$] the $\tau$ polarization $P_\tau$,
   \item[$\bullet$] the effective weak mixing angles $\sin^2 \theta_{eff}^f$.
  \end{itemize}
\end{itemize}
Together with the quantity $\Delta r$ in the correlation of the $W$ mass to the
electroweak input parameters $G_\mu$, $M_Z$ and $\alpha_{EM}$, this set of
precision observables is convenient for a numerical analysis of the
supersymmetric parameter space.

\subsection{The effective $Z$-$f$-$f$ couplings}

The coupling of the $Z$ boson to fermions $f$ can be expressed by effective
vector and axial vector coupling constants $v_{eff}^f, \, a_{eff}^f$ in terms
of the
NC vertex:
\begin{equation}
J_{NC}^\mu = \frac{e}{2 s_W^2 c_W^2} \, \gamma^\mu \, ( v_{eff}^f - a_{eff}^f
\gamma_5 ) \ ,
\end{equation}
where the convention is introduced : $c_W^2 = \cos^2 \theta_W = 1 - s_W^2 =
M_W^2 /
M_Z^2$ \cite{sirlin}.
Input parameters are the $\mu$ decay constant $G_\mu = 1.166392 \times 10^{-5}$
GeV$^{-2}$, $\alpha_{EM} = 1/137.036$ and the mass of the $Z^0$ boson $M_Z =
91.1887$
GeV. The mass of the $W$ boson is related to these input parameters through:
\begin{eqnarray}\label{gmudr}
{G_{\mu}\over\sqrt{2}} = {\pi\alpha_{EM}\over 2 s^2_W M^2_W} \cdot
\frac{1}{
1 - \Delta r_{MSSM} \left(\alpha_{EM},M_W,M_Z,m_t,...\right)} \ ,
\end{eqnarray}
where the complete MSSM one-loop contributions are parametrized by the
quantity $\Delta r_{MSSM}$\cite{sola}.
Leading higher order Standard Model corrections\cite{bhp,fleischer} to the
quantity $\Delta r$ are included in the calculation.
\smallskip

The effective couplings $v_{eff}^f, \, a_{eff}^f$ can be written as:
\begin{eqnarray}
v_{eff}^f & = & \sqrt{Z_Z} \, (v^f + \Delta v^f + Z_M Q_f) \nonumber \\
a_{eff}^f & = & \sqrt{Z_Z} \, (a^f + \Delta a^f) \ .
\end{eqnarray}
$v^f$ and $a^f$ are the tree-level vector and axial vector couplings:
\begin{equation}
v^f = I_3^f - 2 Q_f s_W^2 \ , \ a^f =   I_3^f.
\end{equation}
$Z_Z$, $Z_M$ are given in Eq. (\ref{ZZ}).
The complete MSSM one-loop contributions of the non-universal finite vector and
axial vector couplings $\Delta v^f$, $\Delta a^f$ have been
calculated\cite{dab/hollik},
together with the leading two-loop Standard Model
contributions\cite{bhp,Chet/Kw}.
They are derived in the 't Hooft-Feynman gauge and in the
on-shell renormalization scheme\cite{bhs}.
\smallskip

The universal finite $Z$ boson wave function renormalization $Z_Z$ and the
$\gamma Z$
mixing $Z_M$ are derived from the $(\gamma, \, Z)$ propagator matrix. The
inverse matrix is:
\begin{equation}
(\Delta_{\mu \nu})^{-1} = i g_{\mu \nu} \, \left(
 \begin{array}{ll}
 k^2 + \hat{\Sigma}_\gamma (k^2) &  \hat{\Sigma}_{\gamma Z} (k^2) \\
 \hat{\Sigma}_{\gamma Z} (k^2)   &  k^2 - M_Z^2 + \hat{\Sigma}_Z (k^2)
 \end{array} \right) \ ,
\label{gzmatrix}
\end{equation}
where $\hat{\Sigma}_\gamma$, $\hat{\Sigma}_{Z}$, $\hat{\Sigma}_{\gamma Z}$
are the renormalized self energies and mixing.
They are obtained by summing the loop diagrams and the counter terms\cite{bhp}.
\smallskip

The entries in the $(\gamma, \, Z)$ propagator matrix:
\begin{equation}
\Delta_{\mu \nu} = - i g_{\mu \nu} \left( \begin{array}{ll}
\Delta_\gamma     &  \Delta_{\gamma Z} \\
\Delta_{\gamma Z} &  \Delta_Z \end{array} \right) \ ,
\end{equation}
are given by:
\begin{eqnarray}
\Delta_\gamma (k^2) & = & \frac{1}{k^2 +  \hat{ \Sigma}_\gamma (k^2) -
 \frac{\hat{ \Sigma}_{\gamma Z}^2 (k^2)}{k^2 - M_Z^2 + \hat{ \Sigma}_Z (k^2)} }
 \nonumber \\
 \Delta_Z (k^2) & = & \frac{1}{k^2 - M_Z^2 + \hat{ \Sigma}_Z (k^2) -
 \frac{\hat{ \Sigma}_{\gamma Z}^2 (k^2)}{k^2  + \hat{ \Sigma}_\gamma (k^2)} }
 \nonumber \\
\Delta_{\gamma Z} (k^2) & = & -\frac{\hat{ \Sigma}_{\gamma Z} (k^2)}{
 [ k^2  + \hat{ \Sigma}_\gamma (k^2)] \,
 [ k^2 - M_Z^2 + \hat{ \Sigma}_Z (k^2) ] - \hat{ \Sigma}_{\gamma Z}^2 (k^2) } \
{}.
\label{zprop}
\end{eqnarray}
The renormalization condition to define the mass of the $Z$ boson is given
by the pole of the propagator matrix Eq. (\ref{gzmatrix}).
The pole $k^2 = M_Z^2$ is the solution of the equation:
\begin{equation}
 \cal R\rm e \, [ \, ( M_Z^2   + \hat{ \Sigma}_\gamma (M_Z^2) \, ) \,
 \hat{ \Sigma}_Z (k^2)  - \hat{ \Sigma}_{\gamma Z}^2 (M_Z^2) \, ] = 0 \ .
\end{equation}
Eq. (\ref{zprop}) yields the wave function renormalization $Z_Z$ and mixing
$Z_M$:
\begin{eqnarray}
Z_Z & = & Res_{M_Z} \Delta_Z = \left. \frac{1}{1 + \hat{ \Sigma}_Z' (k^2) -
   \left( \frac{ \hat{ \Sigma}_{\gamma Z}^2 (k^2)}{k^2 + \hat{ \Sigma}_\gamma
(k^2)
   } \right)  } \ \right| _{k^2 = M_Z^2} \nonumber \\
Z_M & = & - \frac{\hat{ \Sigma}_{\gamma Z} (M_Z^2)}{M_Z^2 +
 \hat{ \Sigma}_\gamma (M_Z^2) } \ .
\label{ZZ}
\end{eqnarray}

\subsection{ $Z$ boson observables}

\noindent
The fermionic $Z$ boson partial decay widths $\Gamma_{f \bar{f}}$ can be
written:
\begin{itemize}
\item[1)] $f \ne b$:
\begin{eqnarray}
 \Gamma_{f \bar{f}} & = & \frac{N_C \, \alpha_{EM} \, M_Z}{3}  \,
 \sqrt{ 1 - \frac{4 m_f^2}{M_Z^2} } \, \left[ (v_{eff}^f)^2 (1 + \frac{2 m_f^2}
 {M_Z^2} ) + (a_{eff}^f)^2 (1 - \frac{4 m_f^2}{M_Z^2} ) \right] \cdot \nonumber
\\
  & & \cdot (1 + \frac{3 \alpha_{EM}}{4 \pi} Q_f^2) \, (1 + \delta_{QCD}^f ) \
,
\end{eqnarray}
where
\begin{equation}
 \delta_{QCD}^f = \left\{ \begin{array}{ll}
  0 & ,f = \mbox{leptons} \nonumber \\
  \frac{\alpha_s}{\pi} + 1.405 (\frac{\alpha_s}{\pi})^2 - 12.8
  (\frac{\alpha_s}{\pi})^3 & ,f = \mbox{quarks}
 \end{array} \right. \ .
\label{deqc}
\end{equation}
\item[2)] $f = b$:
\begin{eqnarray}
 \Gamma_{b \bar{b}} & = & \alpha_{EM} \, M_Z  \,
 \left[ (v_{eff}^b)^2 + (a_{eff}^b)^2  \right]
 \cdot (1 + \frac{3 \alpha_{EM}}{4 \pi} Q_b^2) \, (1 + \delta_{QCD}^b ) \,
 + \Delta \Gamma_{b \bar{b}} \ . \nonumber \\
\end{eqnarray}
In $\Delta \Gamma_{b \bar{b}}$ the $b$ quark specific finite mass terms and QCD
corrections together with the leading two-loop $\cal O \rm ( \alpha_{EM}
\alpha_S )$
are included\cite{bhp,Chet/Kw}. $\delta_{QCD}^b$ is given in Eq. (\ref{deqc}).
\end{itemize}
The total decay width $\Gamma_Z$ is the sum of the leptons and quarks: \\
 \begin{equation}
   \Gamma_Z = \sum_{f} \Gamma_{f \bar{f}} \ .
 \end{equation}
 In the following $\Gamma_{had} = \sum_{q} \Gamma_{q \bar{q}}$ is the hadronic
decay
 width of the $Z$ boson.
\medskip

\noindent
The hadronic peak cross section:
\begin{equation}
 \sigma_h = \frac{12 \pi}{M_Z^2} \frac{\Gamma_{ee} \Gamma_{had}}{\Gamma_Z^2} \
{}.
\end{equation}
The ratio of the hadronic to the electronic decay width is defined:
\begin{equation}
 R_e = \frac{\Gamma_{had}}{\Gamma_{e e}} \ .
\end{equation}
The ratio of the partial decay width for $Z \rightarrow b \bar{b} \, (c
\bar{c})$ to
the total hadronic decay width:
\begin{equation}
 R_{b (c)} = \frac{\Gamma_{b \bar{b} (c \bar{c}) }}{\Gamma_{had}} \ .
\end{equation}
\medskip

The following quantities and observables depend on the ratio of the vector to
the axial vector coupling. The effective flavour dependent weak mixing angle
can
be written:
\begin{equation}
 \sin^2 \theta_{eff}^f = \frac{1}{4 | Q_f |} \, \left( 1 -
\frac{v_{eff}^f}{a_{eff}^f}
 \right) \ .
\end{equation}
The \it left-right \rm asymmetries:
\begin{equation}
 A_{LR}^f = {\cal A}^{\rm f}  = \frac{2 \, v_{eff}^f / a_{eff}^f}{1 + (
v_{eff}^f /
 a_{eff}^f )^2 } \ .
\end{equation}
The \it forward-backward \rm asymmetries:
\begin{equation}
 A_{FB}^f = \frac{3}{4} \, \cal A^{\rm e} \rm \, \cal A^{\rm f} \rm \ .
\end{equation}

\section{Discussion}

The experimental results from the recent LEP and $p \bar{p}$ data are
summarized in table \ref{tab1}\cite{Lep1,tevatron}. In addition to the
measurements,
the correlation matrix\cite{Lep1} is used.
The value for $\alpha_S (M_Z^2)$ has been fixed from QCD observables at the Z
pole\cite{QCD}:
\begin{equation}
 \alpha_S (M_Z^2) = 0.123 \pm 0.006 \ ,
\end{equation}
and the $b$ quark mass $m_b = 4.7$ GeV.
In the following we discuss the general version of the MSSM without further
restrictions to the parameters from assumptions about Grand Unification or
Supergravity.

\begin{table}
\begin{center}
\caption{Experimental LEP and  $p \bar{p}$  results.}
\begin{tabular}[t]{||l|rcl||}
\hline
\hline
$\Gamma_Z$ [GeV] & $2.4971$ & $\pm$ & $0.0032$    \\
$\sigma_h$ [GeV] & $41.492$ & $\pm$ & $0.081$     \\
\hline
$R_{  e}$        & $20.843$ & $\pm$ & $0.060$     \\
$R_{  \mu} $     & $20.805$ & $\pm$ & $0.048$     \\
$R_{  \tau}$     & $20.798$ & $\pm$ & $0.066$     \\
$A_{FB}^{ e}$     & $0.0154$ & $\pm$ & $0.0030$    \\
$A_{FB}^{\mu}$    & $0.0160$ & $\pm$ & $0.0017$    \\
$A_{FB}^{\tau}$   & $0.0209$ & $\pm$ & $0.0024$    \\
\hline
$R_b$            & $0.2204$ & $\pm$ & $0.0020$    \\
$R_c$            & $0.1606$ & $\pm$ & $0.0095$    \\
$A_{FB}^b$       & $0.1015$ & $\pm$ & $0.0036$    \\
$A_{FB}^c$       & $0.0760$ & $\pm$ & $0.0089$    \\
\hline
$\Delta r$ & $0.0425$ & $\pm$ & $0.009$     \\
\hline
\hline
\end{tabular}
\label{tab1}
\end{center}
\end{table}

\subsection{Higgs sector}

The Higgs sector of the MSSM is that of a two Higgs doublet model, where the
coefficients of the potential are restricted by supersymmetry. As a consequence
of the supersymmetric Higgs potential, a light Higgs boson exists with a tree
level upper mass bound given by the $Z$ boson mass. The Higgs potential
contains
two independent free parameters, which are $\tan \beta = v_2 / v_1$ and the
mass
of the pseudoscalar Higgs boson $M_{A^0}$. $v_1$, $v_2$ are the vacuum
expectation
values of the Higgs doublets.
\smallskip

Radiative corrections to the Higgs mass spectrum predict an upper limit of the
light Higgs mass $\cal O \rm (130$ GeV$)$\cite{hempf}. In the calculation of
the
neutral scalar MSSM Higgs mass spectrum and the mixing angle $\alpha$ the
corrections
$\cal O \rm (m_t^4)$ are included\cite{dabx}.
\smallskip

Fig. 1 shows the dependence of a subset of observables: $\Delta r$, $R_b$,
$R_c$,
$R_h = R_e$, $\Gamma_{tot}$, $\Gamma_{e e}$, $\sin^2 \theta_{eff}^e$,
$A_{FB}^b$
and $A_{FB}^c$ as functions on $M_A$ for values $\tan \beta = 0.7$ (dotted),
$1.5$ (long dotted), $8$ (dashed-dotted), $20$ (dashed), $70$ (solid).
This set of observables is convenient for a qualitative discussion of the
universal and non-universal MSSM one-loop contributions. The top quark mass
is fixed at $m_t = 174$ GeV, and squark, slepton masses are $m_{\tilde{q}} =
500$ GeV,
$m_{\tilde{l}} = 800$ GeV without left-right mixing. The soft breaking
parameters
in the gaugino sector are $\mu = 100$ GeV, $M = 300$ GeV and a heavy gluino
mass
$m_{\tilde{gl}} = 800$ GeV is fixed. This set of genuine supersymmetric masses
does
not show sizeable effects on these observables and allows to explore only the
SUSY Higgs sector.
\smallskip

A complete discussion on the quantity $\Delta r$ $( \delta M_W )$ is
available\cite{sola}, however, the inclusion of $\cal O \rm (m_t^4)$ Higgs mass
and
$\cal O \rm (\alpha_{EM} \alpha_S^2)$ contributions\cite{dabx,bhp} improve the
theoretical $\Delta r$ predictions.
In Fig. 1 the quantity $\Delta r$ increases with $M_A$ and is
almost constant above $M_A \ge 250$ GeV. Small $\tan \beta = 0.7$ values give
sizeable positive contributions to $\Delta r$, since radiative corrections to
the scalar Higgs mass are large in the regime of small $M_A < 100$ GeV and
small $\tan \beta$. In the
mass range $M_A < 70$ GeV and $\tan \beta \ge 2$ the quantity $\Delta r$
decreases
and disfavours with the experimental data within the range of the measured top
quark mass. Contributions of genuine supersymmetric masses (next subsection) do
not increase $\Delta r$ and enhance the deviation from the experimental result
even stronger.
A similar conclusion for the parameters $M_A$, $\tan \beta$ is obtained
in Fig. 1 for the quantity $\sin^2 \theta_{eff}^e$.
\smallskip

In the $\tan \beta < 1$ range the ratio $R_b$ decreases for values $M_A < 500$
GeV.
This effect is strong for small $M_A \approx 50$ GeV and yields the
Standard Model results for $M_A \rightarrow \infty$. For intermediate $\tan
\beta$,
$4 < \tan \beta < 30$, $R_b$ is nearly constant for $50$ GeV $ \le M_A
\le 500$ GeV and agrees with the Standard Model prediction.
Large $\tan \beta \approx 70$ enhance $R_b$ for small $M_A \le 55$ GeV, within
the experimental $1 \sigma$ bounds. In Fig. 1 this peak at $M_A = 46$ GeV is
shown, however this scenario would predict a rather light neutral Higgs mass
$M_{h^0}$. For $M_A \ge 75$ GeV the ratio $R_b$ decreases, reaches a minimum
for
$M_A = 150$ GeV and approaches the SM result for larger pseudoscalar masses.
The ratio $R_c$ in Fig. 1 lies above the experimental $1 \sigma$ bounds.
Recently,
the complementary $R_b - R_c$ effects have also been discussed\cite{Garcia1}.
However,
the $R_c$
sensitivity on the Higgs sector is small and only for large $\tan \beta
= 70$ and $40$ GeV $< M_A < 60$ GeV the agreement with the experimental data is
improved.
\smallskip

The total decay width $\Gamma_{tot}$ and the ratio $R_h$ in Fig. 1 show a
qualitatively similar line shape as the ratio $R_b$. $\Gamma_{tot}$ and  $R_h$
depend strongly on $\alpha_s$, however, the ``peak'' region in $R_b$ for $M_A <
70$ GeV and small (large) $\tan \beta < 1 ( > 40)$ yields values for the
quantities $\Gamma_{tot}$, $R_h$ which are larger than the experimental
results.
A better agreement can be obtained by decreasing the total value of $\alpha_s$.

\subsection{Sfermion and Gaugino/Higgsino sector}

Sfermions and charginos (neutralinos) appear in the vertex correction to
$Z$ boson decays and give sizeable contributions for the external $b$ state.
The chargino (neutralino) masses and the mixing angles in the gaugino
couplings are calculated from soft breaking parameters $M$, $M'$ and $\mu$ in
the chargino (neutralino) mass matrix\cite{hunter}.
The validity of the GUT relation $M' = 5/3 \tan \theta_W \, M$ is assumed.
\smallskip

Squarks and sleptons are described by a $2 \times 2$ mass matrix:
\begin{equation}
 \cal M_{\rm \tilde{f}}^{\rm 2} \rm = \left( \begin{array}{ll}
 M_{\tilde{Q}}^2 + m_f^2 + M_Z^2 (I_3^f - Q_f s_W^2) \cos 2 \beta &
 m_f (A_f + \mu \{ \cot \beta , \tan \beta \} ) \\
 m_f (A_f + \mu \{ \cot \beta , \tan \beta \} ) &
 M_{\{\tilde{U},\tilde{D}\}}^2 + m_f^2 + M_Z^2 Q_f s_W^2 \cos 2 \beta
 \end{array} \right) \ ,
\label{sqmatrix}
\end{equation}
with SUSY soft breaking parameters $M_{\tilde{Q}}$,
$M_{\tilde{U}}$, $M_{\tilde{D}}$, $A_f$, and $\mu$. The notation in the
off-diagonal entries in Eq. (\ref{sqmatrix}):
\begin{equation}
A_f' = A_f + \mu \{ \cot \beta , \tan \beta \} \ ,
\label{glaprime}
\end{equation}
will be used. Up and down type sferminos in (\ref{sqmatrix}) are distinguished
by
setting f=u,d and the $\{u,d\}$ entries in the parenthesis.
Instead of $M_{\tilde{Q}}$, $M_{\tilde{U}}$, $M_{\tilde{D}}$, $A_b'$, $A_t'$,
the physical squark masses are given by
$m_{\tilde{b}} = m_{\tilde{b}_L} = m_{\tilde{b}_R}$, $m_{\tilde{t}_2}$ and
$A_t'$. No left-right mixing for $\tilde{b}$ squarks is assumed and the
$\tilde{u}$, $\tilde{d}$, $\tilde{c}$, $\tilde{s}$ masses are equal to the
$\tilde{b}$ squark mass. The sneutrino masses  $m_{\tilde{\nu}}$ are given
without
left-right mixing for sleptons.
\smallskip

In Fig. 2 the dependence of the $Z$ boson observables and $\Delta r$ for
sbottom masses $m_{\tilde{b}} = 150 \, (500)$ GeV, and stop masses
$m_{\tilde{t}_R} = 50$ GeV (solid line), $150$ GeV (dotted, dotted-dashed),
$300 \, (500)$ GeV (long dotted, dashed) is shown as a function of $\tan
\beta$.
No left-right mixing in the stop states is assumed. $m_t = 174$ GeV,
$\mu = 100$ GeV, $M = 100$ GeV, $M_A = 800$ GeV, $m_{\tilde{l}} = 900$ GeV and
$m_{\tilde{gl}} = 900$ GeV.
\smallskip

Sfermion and chargino (neutralino) contributions decrease the quantity
$\Delta r$\cite{sola}. Charginos and neutralinos give numerically small
effects,
large $\tilde{u}_L$-$\tilde{d}_L$ mass splittings in the squark and slepton
sector are the dominiant supersymmetric contributions.
The quantity $\Delta r$ gets sizeable contributions from the $\rho$ parameter:
\begin{equation}
 \Delta r \simeq \Delta \alpha_{EM} - \frac{c_W^2}{s_W^2} \, \Delta \rho \, +
... \ .
\end{equation}
The $\tilde{b}_L$-$\tilde{t}_L$ mass splitting is described by the $\Delta
\rho$
parameter\cite{sola}:
\begin{equation}
\Delta \rho_{\tilde{b}-\tilde{t}}^0 = \frac{3 \alpha_{EM}}{16 \pi s_W^2 M_W^2}
\,
 \left(m_{\tilde{b}_L}^2 + m_{\tilde{t}_L}^2 - 2 \frac{m_{\tilde{b}_L}^2
 m_{\tilde{t}_L}^2}{m_{\tilde{b}_L}^2 - m_{\tilde{t}_L}^2} \log \frac{
 m_{\tilde{b}_L}^2}{m_{\tilde{t}_L}^2} \right) \ ,
\end{equation}
and decreases $\Delta r$ for light $\tilde{b}$ masses\footnote{The superscript
$\Delta \rho^0$ indicates that no left-right mixing is present.}.
Fig.\,2 shows $\Delta r$ for sbottom masses $m_{\tilde{b}} = 150 \, (500)$ GeV,
where a light $\tilde{b}$ mass decreases $\Delta r$ sizeably. A variation in
the
stop state masses $m_{\tilde{t}_R}$ in Fig.\,2 shows small contributions on
$\Delta r$. The effect on left-right sfermion mixing is discussed in Fig.\,3.

An enhancement of the ratio $R_b$ can be observed for light charginos $\cal O
\rm
( 100$ GeV$)$ and light stop-sbottom masses. This effect is stronger for
small (large) $\tan \beta < 1 ( > 40) $ values due to the $\tan \beta$ enhanced
Yukawa type chargino (neutralino) - quark - squark couplings.
Taking the present experimental limits on the genuine supersymmetric and Higgs
masses
into account, $R_b$ can not be increased within the experimental $1 \sigma$
bound.
The enhancement in $R_b$ is larger for light
$\tilde{t}_R$ masses in Fig. 2, whereas $R_b$ is nearly insensitive to
the sbottom mass and therefore the $\tilde{t}_L$ mass.
\smallskip

The ratio $R_h$ and the total decay width $\Gamma_{tot}$ increase for light
left or right sfermion states in Fig. 2. For $m_t = 174$ GeV and $\alpha_s =
0.123$,
a light $\tilde{b}$ squark $m_{\tilde{b}} = 150$ GeV increases $\Gamma_{tot}$
by about
$2 \sigma$ above the present experimental data. This enhancement for
light sbottom masses is shown for the electronic decay width $\Gamma_{ee}$,
where
no sensitivity for $\tilde{t}_R$ masses can be observered.
The \it forward-backward \rm  asymmetry $A_{FB}^b$ also favours larger
sbottom masses and shows smaller effects from $m_{\tilde{t}_R}$. Light
$\tilde{b}_L$
masses increase the \it forward-backward \rm asymmetries $A_{FB}^b$,
$A_{FB}^c$.
\smallskip

Sfermion mixing for the parameter set in Fig.\,2 ``screens'' the contributions
from
supersymmetric particles for all displayed observables.
In Fig.\,3,  $m_{\tilde{b}} = 150$ GeV, $m_{\tilde{t}_2} = 50$ GeV and all
other
parameters are as in Fig.\,2. The stop mixing is shown for
$A_t' = -930$ GeV (dotted-dashed), $-464$ GeV (dashed), $0$ (solid), $464$ GeV
(dotted), $930$ GeV (long dotted). These $A_t'$ mixing values yield
$\tilde{t}_1$
masses $m_{\tilde{t}_1} = 800$, $442$, $216$, $442$, $800$ GeV.
No (very small) stop mixing decreases $\Delta r$ maximal, increases the ratio
$R_b$
maximal, etc.
As a result, sfermion mixing can screen the supersymmetric radiative
corrections
for precision observables,  even if supersymmetric particles are light
$\cal O \rm (100$ GeV$)$.
\smallskip

Fig.\,4 shows the $Z$ boson observables and $\Delta r$ for chargino
(neutralino) masses as functions of $\mu$ for fixed $M = 100$ GeV. Stop masses
are $m_{\tilde{t}_R} = 50$ GeV (solid line), $150$ GeV (dotted), $800$ GeV
(long dotted) and $m_{\tilde{b}} = 500$ GeV without left-right mixing.
$m_t = 174$ GeV and  large $A^0$, slepton and gluino masses are assumed:
$M_A = 800$ GeV, $m_{\tilde{l}} = 900$ GeV, $m_{\tilde{gl}} = 800$ GeV.
In Fig. 4 $\tan \beta = 1.1$ and the experimentally
excluded chargino masses are indicated by the shadowed $\mu$ range
($m_{\tilde{\chi}^\pm} \ge 48$ GeV). Close to the experimentally exluded $\mu$
region
the contribution from charginos and sfermions is largest. The ratio $R_b$
increases
within the experimental $1 \sigma$ range for $m_{\tilde{\chi}^\pm} = 50$ GeV
and $m_{\tilde{t}_R} = 50$ GeV. Larger chargino/stop masses decouple above
$\cal O \rm (500$ GeV$)$ from the discussed observables.
Larger $\tan \beta$ values give a smaller but qualitative similar result for
$R_b$.
A small value $\tan \beta = 1.1$ decreases $\Delta r$ also for light
$\tilde{t}_R$
masses. The ratio $R_h$, $\sin^2 \theta_{eff}$ and the \it forward-backward \rm
asymmetry $A_{FB}^b$ are in better agreement with the experimental data for
large
chargino/stop masses or strong sfermion mixing.

\subsection{Upper and lower bounds at LEP I and LEP II}

Fig. 5 shows the overall upper and lower bounds of the theoretical predictions
for each observable individually.
The results are plotted as a function of the top quark mass $130$ GeV $\le m_t
\le
200$ GeV for the Standard
Model (dashed) and the MSSM (solid). The standard Higgs mass is varied between
$66$ GeV $\le M_{H_{SM}} \le 800$ GeV. The MSSM parameter set is restricted by
the present mass bounds from direct searches at \it LEP I \rm and $p \bar{p}$
data as shown by the solid MSSM lines in Fig. 5. Mass bounds from direct
sparticle
searches to be expected
at \it LEP~II \rm are shown by dotted lines. An overlap between the
SM and MSSM upper and lower bounds is given for large decoupled genuine
supersymmetric particles and a light neutral scalar MSSM Higgs mass:
$M_{h^0_{upper}} > M_{H_{SM, lower}}$.
The experimental results are indicated by the dark bounds.
\smallskip

In case that no direct signal for supersymmetry is seen at \it LEP II \rm,
upper and lower MSSM bounds on $R_b$ ($R_c$) can not explain the presently
observed discrepancy with the experimental data.
In conclusion from the previous results and Fig. 5 the
MSSM can not improve the predictions for precision observables for light
sfermions $\le 300$ GeV. Constraints on sfermion masses for top quark masses
$160$ GeV $\le m_t \le 200$ GeV and $\alpha_s = 0.118 ... 0.128$ prefer
squarks $ m_{\tilde{q}} > 300$ GeV, if no squark mixing is assumed.
Large sfermion mixings, however, can not exclude a light stop quark
$\cal O \rm (100$ GeV$)$.

\section{Acknowledgements}

We thank B. Kniehl and K. Nathan for the excellent organization during the
Ringberg Workshop.

\newpage

\section{Figure Captions}

\noindent{\bf Figure 1.}~
The observables $\Delta r$, $R_b$, $R_c$, $\Gamma_{tot}$, $\Gamma_{ee}$,
$\sin^2 \theta_{eff}^e$, $A_{FB}^b$, $A_{FB}^c$ are plotted as a function
on thepseudoscalar mass $M_A$ for values $\tan \beta = 0.7$ (dotted line),
$1.5$ (long dotted), $8$ (dashed-dotted), $20$ (dashed) and $70$ (solid).
$m_t = 174$ GeV, squark and slepton masses are $m_{\tilde{q}} = 500$ GeV,
$m_{\tilde{l}} = 800$ GeV, no left-right mixing. $\mu = 100$ GeV,
$M = 300$ GeV and the gluino mass $m_{\tilde{gl}} = 800$ GeV.

\vskip 0.3cm

\noindent{\bf Figure 2.}~
The set of observables from Fig. 1. as a function on $\tan \beta$ for
squark masses $m_{\tilde{b}} = 150 \, (500)$ GeV and
$m_{\tilde{t}_R} = 50$ GeV (solid line), $150$ GeV (dotted), $300 \, (500)$
GeV (long dotted, dashed), no left-right mixing.
$m_t = 174$ GeV, $\mu = 100$ GeV, $M = 100$ GeV, $M_A = 800$ GeV,
$m_{\tilde{l}} = 900$ GeV and $m_{\tilde{gl}} = 900$ GeV.

\vskip 0.3cm

\noindent{\bf Figure 3.}~
The set of observables from Fig. 1. as a function on $\tan \beta$ for
stop mixing values $A_{\tilde{t}'} = -930$ GeV (dotted-dashed), $-464$ GeV
(dashed), $0$ (solid), $464$ GeV (dotted), $930$ GeV (long dotted).
$m_{\tilde{b}} = 150$ GeV, $m_{\tilde{t}_2} = 50$ GeV and all other
parameters from Fig. 2.

\vskip 0.3cm

\noindent{\bf Figure 4.}~
The set of observables from Fig. 1. as a function on $\mu$ for fixed
$M = 100$ GeV. Stop masses are $m_{\tilde{t}_R} = 50$ GeV (solid line),
$150$ GeV (dotted) and $800$ GeV (long dotted). $m_{\tilde{b}} = 500$ GeV,
no left-right mixing, $m_t = 174$ GeV, $M_A = 800$ GeV,
$m_{\tilde{l}} = 900$ GeV, $m_{\tilde{gl}} = 800$ GeV.

\vskip 0.3cm

\noindent{\bf Figure 5.}~
Upper and lower bounds of the individual observables as a function on $m_t$
for the Standard Model (dashed line) and the MSSM (solid line).
The MSSM parameters are restricted by the present mass bounds from
direct searches at $LEP I$ and $p \bar{p}$ colliders. Mass bounds from
direct sparticle searches to be expected at $LEP II$ are shown by the
dotted lines.

\newpage

\begin{figure}
\epsfig{figure=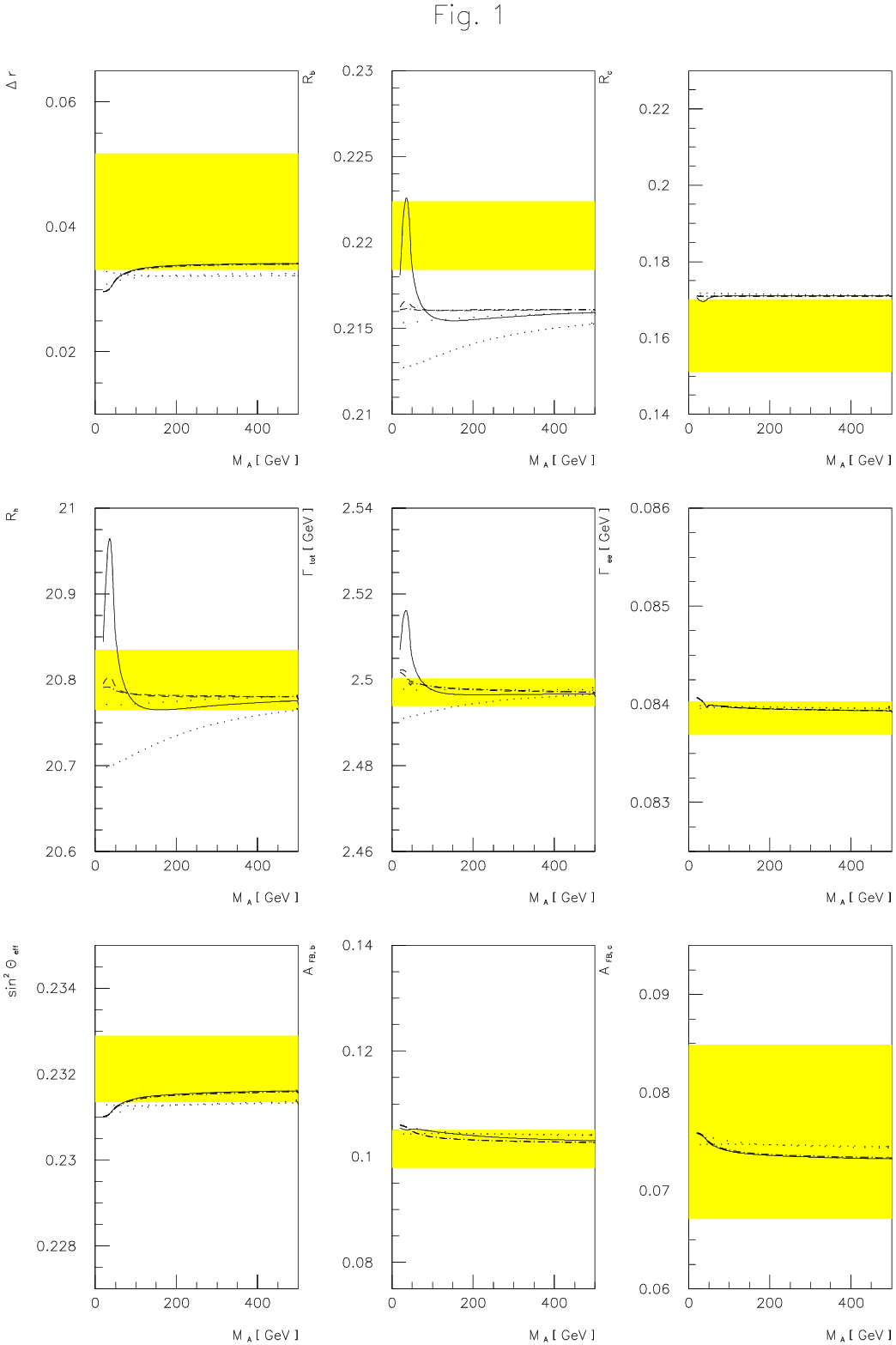,%
height=22cm,width=15.0cm,%
bbllx=75pt,bblly=75pt,bburx=555pt,bbury=751pt}
\end{figure}
\newpage
\begin{figure}
\epsfig{figure=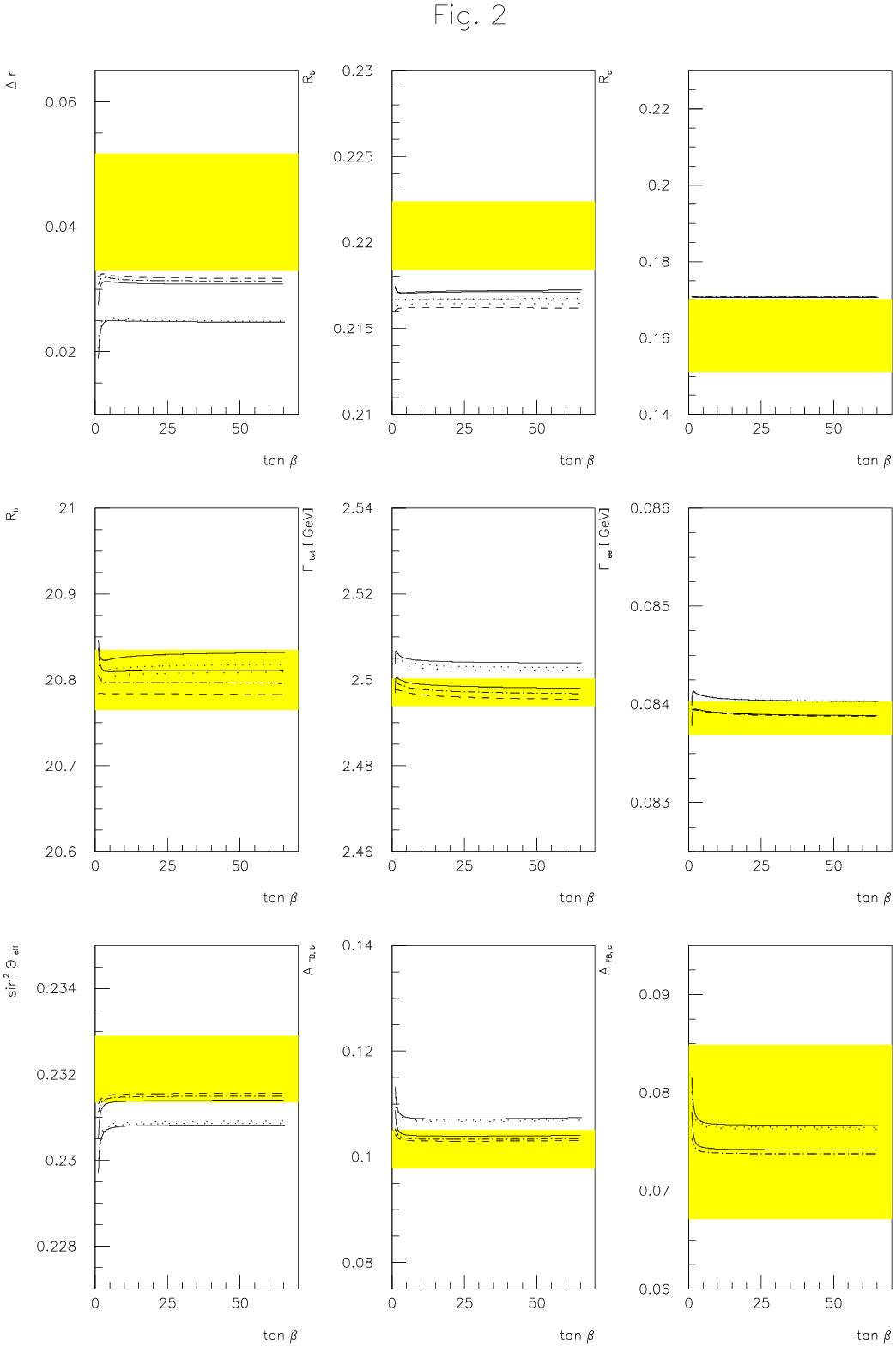,%
height=22cm,width=15.0cm,%
bbllx=75pt,bblly=75pt,bburx=555pt,bbury=751pt}
\end{figure}
\newpage
\begin{figure}
\epsfig{figure=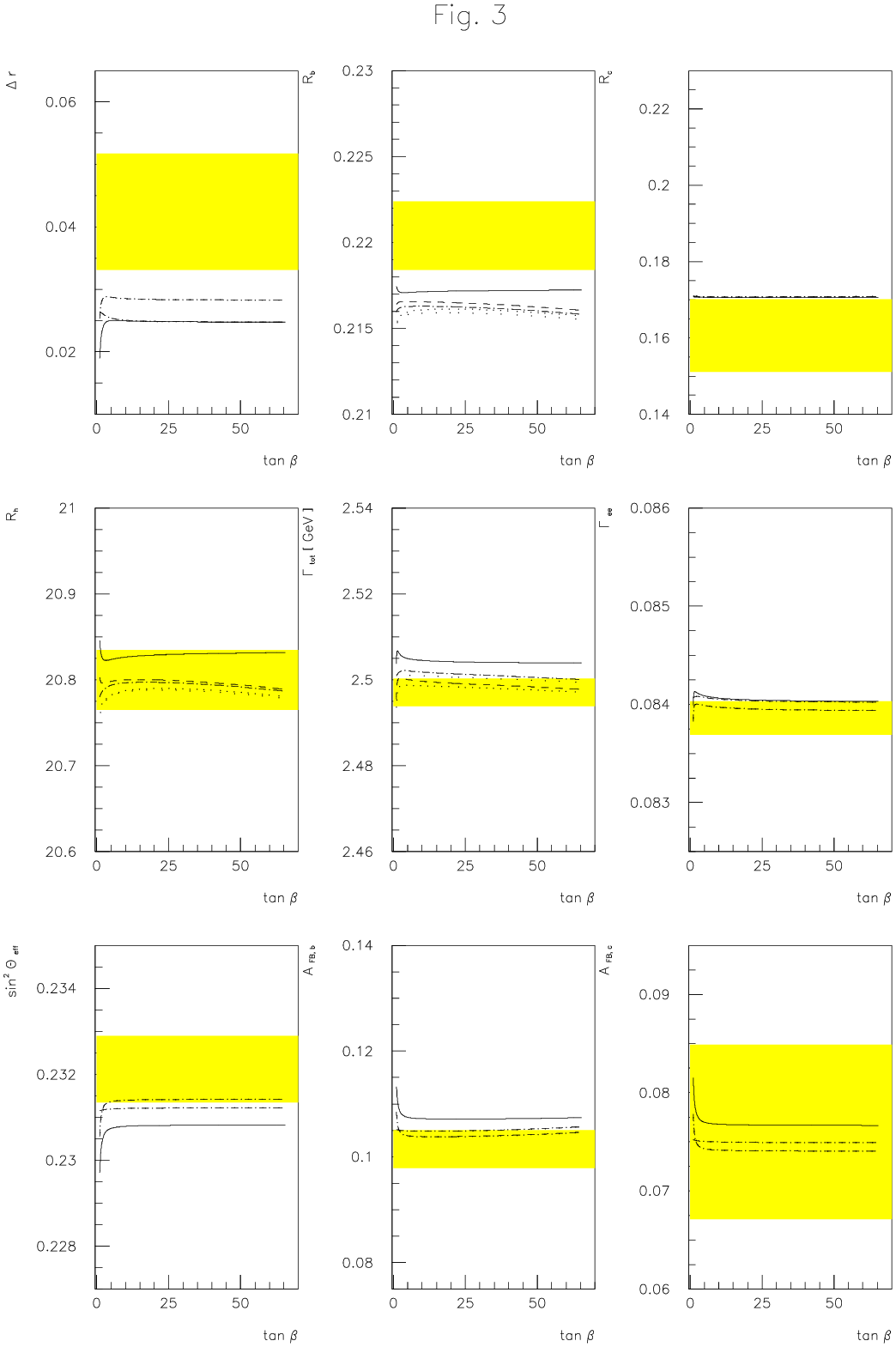,%
height=22cm,width=15.0cm,%
bbllx=75pt,bblly=75pt,bburx=555pt,bbury=751pt}
\end{figure}
\newpage
\begin{figure}
\epsfig{figure=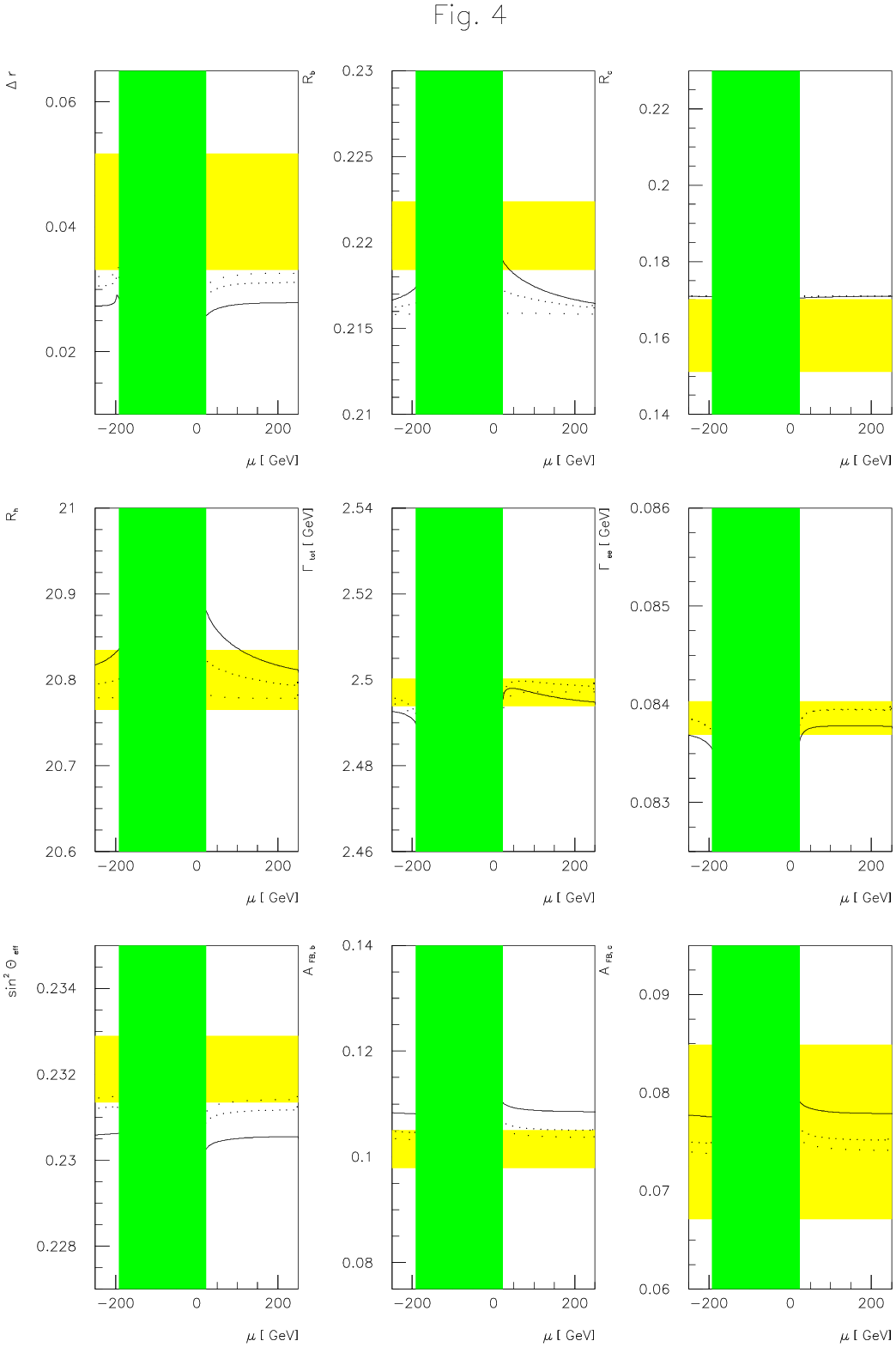,%
height=22cm,width=15.0cm,%
bbllx=75pt,bblly=75pt,bburx=555pt,bbury=751pt}
\end{figure}
\newpage
\begin{figure}
\epsfig{figure=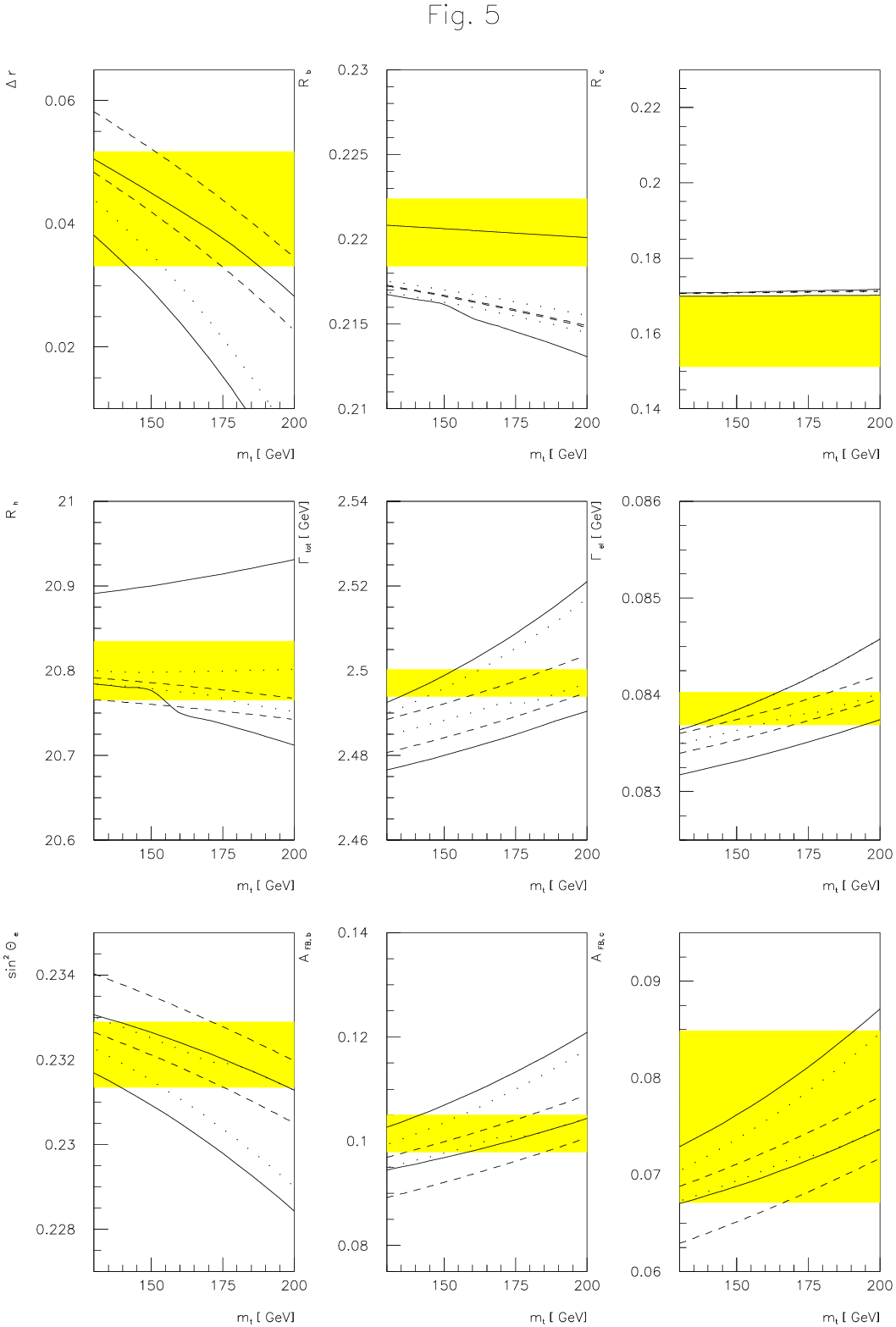,%
height=22cm,width=15.0cm,%
bbllx=75pt,bblly=75pt,bburx=555pt,bbury=751pt}
\end{figure}
\end{document}